\documentstyle[prb,aps,floats,twocolumn,graphicx]{revtex}

\def\sb#1{$_{#1}$}

\begin{document}
\draft

\wideabs{
\title{  Local structure study of $\bf In_{x}Ga_{1-x}As$ semiconductor alloys 
using High Energy Synchrotron X-ray Diffraction }

\author {I.-K. Jeong, F. Mohiuddin-Jacobs, V. Petkov and S.~J.~L. Billinge}

\address{Department of Physics and Astronomy and Center for
         Fundamental Materials Research,\\
         Michigan State University, East Lansing, MI 48824-1116, USA}

\author {S. Kycia}

\address{National Synchrotron Light Laboratory, Sao Paulo, Brazil}

\date{\today}

\maketitle

\begin{abstract}

Nearest and higher neighbor distances as well as bond length
distributions (static and thermal) of the $\rm In_{x}Ga_{1-x}As$ $(0
\leq x \leq 1)$ semiconductor alloys have been obtained from high
real-space resolution atomic pair distribution functions (PDFs).
Using this structural information, we modeled the local atomic displacements
in $\rm In_{x}Ga_{1-x}As$ alloys.
From a supercell model based on the Kirkwood potential, we obtained 
3-D As and (In,Ga) ensemble averaged probability distributions. 
This clearly shows that
As atom displacements are highly directional and
can be represented as a combination of $\langle 100\rangle$ and
$\langle 111\rangle$ displacements. 
Examination of the Kirkwood model indicates that
the standard deviation ($\sigma$) of the static disorder on the (In,Ga) 
sublattice is around 60\% of the value on the As sublattice
and the (In,Ga) atomic displacements are much
more isotropic than those on the As sublattice.
The single crystal diffuse scattering calculated from the Kirkwood model 
shows that atomic displacements are most strongly correlated along 
$\langle 110 \rangle$ directions. 
\end{abstract}
}

\pacs{61.72.Dd,61.43.Dq,61.43.Bn,61.12.Ld}



\section{Introduction}

Semiconductor alloys are known as technologically important materials
for their wide applications in optoelectronic devices such as lasers
and detectors.~\cite{glass;sci87} The local structure
information is of fundamental importance in understanding the alloy
systems because their physical properties are strongly influenced by
the local atomic displacements present in the alloys.  For example, it
is known that chemical and compositional disorder strongly affect
the electronic structure of zinc-blende type
alloys~\cite{zunge;prl83,hass;prl84,hwang;prl88,ling;prb88,ekpen;prb90,li;prb92}
and their enthalpies of formation.~\cite{silve;prb95,bella;prb97} \par
 
In this paper we present a detailed study of the local and average
structure of the $\rm In_xGa_{1-x}As$ alloy series.  The average
structure of $\rm In_xGa_{1-x}As$ was studied by Woolley.~\cite{wooll;bk62}
The structure is of the zinc blende type
($F4\overline 3m$)~\cite{wycko;bk67} over the entire
alloy range.  The lattice parameters, and therefore the average In-As
and Ga-As bond lengths, interpolate linearly between the values of the
end-members according to Vegard's law.~\cite{vegar;zp21}  However,
consideration of the local structure reveals a very different
situation.  The local structure of $\rm In_xGa_{1-x}As$ was first
studied by Mikkelson and Boyce using extended x-ray absorption fine
structure (XAFS).~\cite{mikke;prl82} According to this experiment, the
individual nearest neighbor (NN) Ga-As and In-As distances in the
alloys are rather closer to the pure Ga-As and In-As
distances. Further XAFS experiments showed that this is quite general
behavior for many zinc-blende type alloy
systems.~\cite{balza;prb85,boyce;jcg89,wu;prb93} Since then a number of
theoretical and model studies have been carried out on the
semiconductor alloys to understand how the alloys accommodate the local
displacements.~\cite{marti;prb84,shih;prb85,chen;prb85,schab;prb91,cai;prb92ii,podgo;ssc85,sher;prb87,wu;jpcm94}

Until now these models and theoretical predictions are tested mainly
by the comparison with XAFS data.  The XAFS results give information
about the nearest neighbor and next nearest neighbor distances in the
alloys but imprecise information about bond-length distributions and
no information about higher-neighbor shells.  This limited structural
data makes it difficult to differentiate between competing models for
the local structure.  For example, even a simple radial force
model~\cite{shih;prb85} rather accurately predicts the nearest
neighbor distances of $\rm In_xGa_{1-x}As$ alloys in the dilute limit.
Therefore, one needs more complete structural information including
nearest neighbor, far-neighbor distances, and bond length {\it distributions}
to prove the adequacy of model structures for these
alloys.

The atomic pair distribution function (PDF), $G(r)$, measures the
probability of finding an atom at a distance $r$ from another
atom.~\cite{billi;loc98} One of the advantages of the PDF method over
other local probes such as XAFS is that it gives both local {\it and}
intermediate range information because both Bragg peaks and diffuse
scattering are used in the analysis.  It is also possible to
obtain information about the static bond length distribution from
the PDF peak width and about correlations of atom 
displacements.~\cite{jeong;jpca99} 

In this paper, we present a detailed X-ray diffraction study of $\rm
In_xGa_{1-x}As$, $(0\leq x\leq 1)$.  A preliminary analysis of the 
data has been published elsewhere.~\cite{petko;prl99}
Using high energy synchrotron x-rays, we measured the total scattering 
structure function, $ S(Q)$,
of the $\rm In_xGa_{1-x}As$ alloy system extended to high $Q$
($Q_{max}=45$ \AA$^{-1}$) where $Q$ is the magnitude of the momentum
transfer of the scattered x-rays 
($Q=4\pi\sin\theta/\lambda$ for elastic scattering).
From these structure functions we obtained the corresponding high
real-space resolution PDFs through a Fourier transform according to
\begin{equation}
G(r)={{2\over\pi}\,\int^\infty_0Q[S(Q)-1]\sin\,Qr\,dQ}.
\protect\label{eq;gr}\end{equation} In these PDFs, the first peak is
clearly resolved into two sub-peaks corresponding to the Ga-As and
In-As bond lengths.~\cite{petko;prl99} The evolution of the
bond-length with doping gives good agreement with XAFS.  For the
far-neighbor peaks, the peak widths are much broader in the alloy
samples than those of the pure end-members reflecting the increased
disorder.  We model the local structure of $\rm In_xGa_{1-x}As$ alloys
using a supercell model~\cite{chung;prb97} based on the Kirkwood
potential~\cite{kirkw;jcp39} which gives good agreement with the alloy
data with no adjustable parameters. The results of the modeling have
been analyzed to reveal the average atomic static distribution on the
As and (In,Ga) sublattices.  Finally, we have calculated the diffuse
scattering that one would get from the Kirkwood model. This compares
qualitatively well with published diffuse scattering results from $\rm
In_{0.53}Ga_{0.47}As$.~\cite{glas;pmb90}


\section{Experimental details}
\subsection{Data collection}

The alloy samples, with compositions 
$\rm In_xGa_{1-x}As$, (x = 0, 0.17, 0.33, 0.5, 0.83, 1) were prepared 
by a melt and quench method.  An appropriate
fraction of InAs and GaAs crystals were powdered, mixed and sealed
under vacuum in quartz ampoules.  The samples were heated beyond the
liquidus curve of the respective alloy~\cite{wooll;bk62,cunne;bk62} to
melt them and held in the molten state for 3 hours before quenching
them in cold water. The alloys were powdered, resealed in vacuum, and
annealed just below the solidus temperature for 72-96 hours to
increase the homogeneity of the samples. This was repeated until the
homogeneity of the samples, as tested by x-ray diffraction, was
satisfactory.  X-ray diffraction patterns from all the samples showed
single, sharp diffraction peaks at the positions expected for the
nominal alloy similar to the results obtained by Mikkelson and
Boyce.~\cite{mikke;prl82}

High energy x-ray powder diffraction measurements were conducted at
the A2 wiggler beamline at Cornell High Energy Synchrotron Source
(CHESS) using intense x-rays of 60 KeV ($\lambda=0.206$ \AA). The
incident x-ray energy was selected using a Si(111) double-bounce
monochromator. All measurements were carried out in flat plate
symmetric transmission geometry.  In order to minimize thermal atomic
motion in the samples, and hence increase the sensitivity to static
displacements of atoms, the samples were cooled down to 10~K using a
closed cycle helium refrigerator mounted on the Huber 6 circle
diffractormeter. The samples were uniform flat plates of loosely
packed fine powder suspended between thin foils of kapton tape. 
The sample thicknesses were adjusted to achieve
sample absorption $\mu t\sim 1$ for the 60~KeV x-rays, where $\mu$ is
the linear absorption coefficient of the sample and $t$ is the sample
thickness.

The experimental data were collected up to $ Q_{max}=45$ \AA$^{-1}$
with constant $\Delta Q$ steps of 0.02 \AA$^{-1}$.  
This is a very high momentum transfer for x-ray diffraction measurements.
For comparison, $Q_{max}$ from Cu $K_{\alpha}$ x-ray tube is 
less than 8 \AA$^{-1}$. 
This high $Q_{max}$ is crucial to 
resolve the small difference ($\approx 0.14$ \AA)  in the In-As and Ga-As bond lengths.

To minimize the measuring time, the data were collected in two parts,
one in the low $Q$ region from 1 to 13 \AA$^{-1}$ and the other in
mid-high $Q$ region from 12 to 50 \AA$^{-1}$. Because of the intense
scattering from the Bragg peaks, in the low $Q$ region the incident
beam had to be attenuated using lead tape to avoid detector
saturation.  The maximum intensity was scaled so that the count rate
across the whole detector energy range in the Ge detector did not
exceed $\sim 2\times 10^4~s^{-1}$.  At these count-rates detector
dead-time effects are significant but can be reliably corrected as we
describe below.  To reduce the random noise level below 1\%, we
repeated runs until the total \emph{elastic scattering} counts become
larger than 10,000 counts at each value of $Q$.  Also, to obtain a
better powder average the sample was rocked with an amplitude of $\pm
0.5^{\circ}$ at each $Q$-position.  The scattered x-rays were detected
using an intrinsic Ge solid state detector.  The signal from the Ge
detector was processed in two ways. The signal was fed to a
multi-channel analyzer (MCA) so that a complete energy spectrum was
recorded at each data-point.  The signals from the elastic and Compton
scattered radiation could then be separated using software after the
measurement. In parallel, the data were also fed through
single-channel pulse-height analyzers (SCA) which were preset to
collect the elastic scattering, Compton scattering, and a wider energy
window to collect both the elastic and Compton signals. For
normalization, the incident x-ray intensity was monitored using an ion
chamber detector containing flowing Ar gas.

\begin{figure}[!tb]
 \centering \includegraphics[angle=0,width=3.3in]{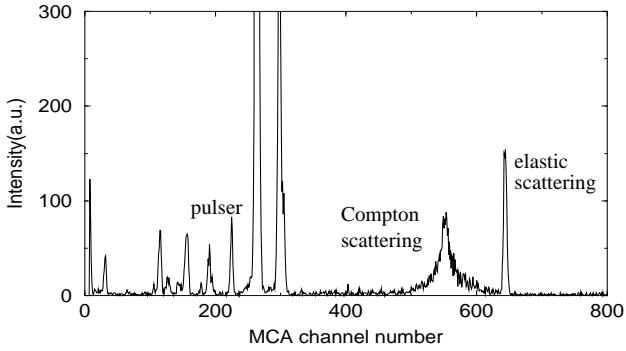}
  \caption{MCA spectrum of $\rm InAs$ at Q=45 \AA$^{-1}$. Peaks in the
  spectrum from the elastic and Compton scattering are labelled, as is
  a peak from an electronic pulser used for dead-time correction.  The
  other peaks in the spectrum come from various fluorescence and
  escape peaks.}  
  \label{fig;fig1}
\end{figure}

For the SCAs, the proper energy channel setting for the elastic
scattering is crucial. Any error in the channel setting could cause an
unknown contamination by Compton scattering and make data corrections
very difficult.  There's no such problem in the MCA method since the
entire energy spectrum of the scattered radiation is measured at each
value of $Q$.  The main disadvantage of the MCA method is that it has
a larger dead-time, although this can be reliably corrected as we show
below.  Fig.~\ref{fig;fig1} shows a representative MCA spectrum
taken from the InAs sample at $Q=45$~\AA . It is clear that the
Compton and elastic scattering are well resolved at this high momentum
transfer.  The elastically scattered signal, which contains the
structural information, is obtained by integrating the area under the
elastic scattering peak. 

\subsection{Data Analysis}

The measured x-ray diffraction intensity may be
expressed~\cite{wased;bk80} by
\begin{equation}
  I^{mea}(Q) = PA[N(I^{coh}_{eu}+I^{inc}_{eu}+I^{mul}_{eu})], 
\end{equation}
where {\em P} is the polarization factor, {\em A} the absorption
factor, {\em N} the normalization constant, and $I^{coh}_{eu},\
I^{inc}_{eu},\ I^{mul}_{eu}$ are the coherent single scattering,
incoherent (Compton), and multiple scattering intensities,
respectively, per atom, in electron units. The total scattering
structure function, $S(Q)$, is then defined as 
\begin{equation}
  S(Q) = [I^{coh}_{eu} - (\langle f^2 \rangle - 
{\langle f \rangle }^2)]/{\langle f \rangle }^2,
  \label{eq;soq}
\end{equation}
where $\langle f \rangle=\langle f(Q) \rangle $ is the sample average
atomic form factor and $\langle f^2 \rangle$ is the sample average of
the square of the atomic form factor. Therefore, to
obtain $S(Q)$ from the measured diffraction data, we have to apply
corrections such as multiple scattering, 
polarization, absorption, Compton scattering and Laue diffuse corrections on
the raw data.~\cite{wased;bk80,wagne;jnon78}

The corrections were carried out using a home-written computer program,
PDFgetX~\cite{jeong;unpub00} that is able to utilize the MCA data.  The
results obtained using the MCA approach are very similar to those obtained
using the SCA approach.~\cite{petko;prl99}  It appears that both approaches
work well for quantitative high energy x-ray powder diffraction.  One possible
advantage of the MCA method is that energy windows of interest can  be set
after the experiment is over which is precluded if data are only collected
using SCA's.

We briefly describe some of the features of the data correction using
PDFgetX.  Data are first corrected for detector dead-time.  In this
experiment, we used the pulser method.~\cite{ander;nucl69} A
pulse-train from an electronic pulser of known frequency is fed into
the detector preamp.  The voltage of the pulser pulses are set so that
the signal appears in a quiet region of the MCA spectrum.  The
measured counts in the pulser signal in the MCA (or, indeed, in an SCA
window set on the pulser signal) is then recorded for each data point.
The data dead-time correction is then obtained by scaling the raw data
by the ratio of the known pulser frequency and the measured pulser
counts.  This method accounts for dead-time in the preamp, amplifier
and MCA/SCA electronics but not in the detector itself.  However, in
general the dead-time is dominated by the pulse-shaping time in the
amplifier or the analogue-digital conversion in the MCA or SCA and so
this method gives rather accurate dynamic measurement of the
detector dead-time.  An alternative dead-time correction protocol for
correcting MCA data is to use the MCA real-time/live-time ratio.  This
works reasonably well if the MCA conversion time is the dominant
contribution to the detection dead-time.  This approach gave similar
results to the pulser correction in this case.

\begin{figure}[!tb]
 \centering \includegraphics[angle=0,width=3.3in]{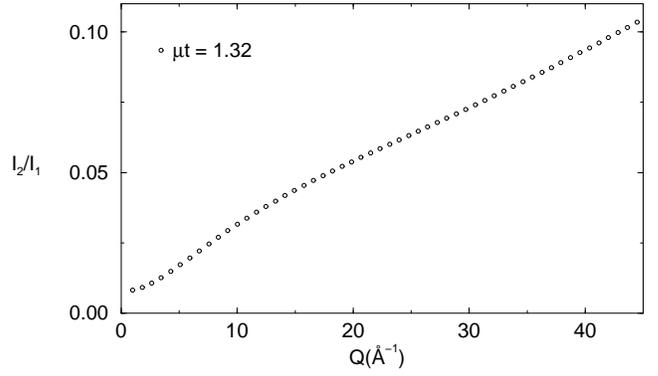}
  \caption{ Calculated double scattering ratio, $I_2/I_1$ where $I_2$ 
   is the intensity due to double scattering events and $I_1$ the single 
   scattering intensity, for the $\rm In_{0.5}Ga_{0.5}As$ alloy 
   in transmission geometry with a $\mu t=1.32$ appropriate for our sample.}
  \label{fig;fig2}
\end{figure}

Multiple (mainly double) scattering can be a problem if samples are
relatively thick and the radiation is highly penetrating as in the
present case.  The multiple scattering contribution contains no usable
structural information and must be removed from the measured
intensity. It depends on sample thickness and many other sample dependent 
factors such as attenuation coefficient, atomic number and weight of sample 
constituent.~\cite{warre;bk90,dwigg;ac71,serim;jac90} 
It increases as the sample becomes thicker in both transmission and reflection
geometry. 
The multiple scattering correction was calculated using the approach
suggested by Warren~\cite{warre;bk90,dwigg;ac71,serim;jac90} in the
isotropic approximation.  Calculation of the multiple scattering intensity is
considerably simplified when the elastic and Compton signals are separated
as is done here since only completely elastic multiple scattering events 
need to be considered.
In
$\rm In_xGa_{1-x}As$ samples, the multiple scattering ratio was around
10$\%$ maximum at high $Q$ in transmission geometry.  This result
suggests that the proper multiple scattering correction becomes
important in the high $Q$ region. Fig.~\ref{fig;fig2} shows the double
scattering ratio calculated for the $\rm In_{0.5}Ga_{0.5}As$ sample.

The x-ray polarization correction is almost negligible for synchrotron x-ray
radiation because the incident beam is almost completely plane polarized
perpendicular to the
scattering plane. As a result there is virtually no angle dependence to the
measured intensity  due to polarization effects.~\cite{warre;bk90}

\begin{figure}[!tb]
  \centering \includegraphics[angle=0,width=3.3in]{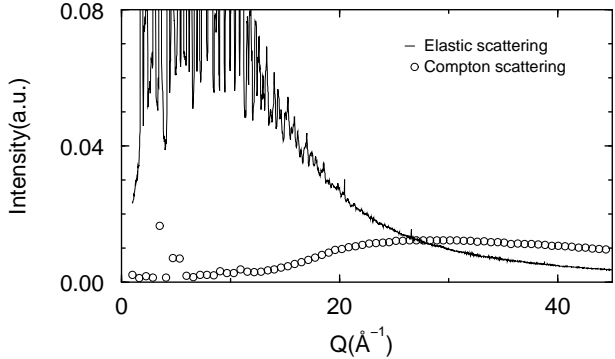}
  \caption{ Comparison between Compton and elastic scattering intensities
   measured in $\rm In_{0.5}Ga_{0.5}As$.}
 \protect
 \label{fig;fig3}
\end{figure}

The Compton scattering correction is very important
in high-energy x-ray diffraction data analysis.  It can become larger
than the coherent scattering intensity at high $Q$, as is evident in
Fig.~\ref{fig;fig1}. Even a small error in determining the Compton
correction can lead to a significant error in the coherent scattering 
intensity in the 
high $Q$ region.  However, in this region of the diffraction pattern
the elastic and Compton-shifted scattering are well separated in
energy and can be reliably separated using the energy resolved detection
scheme we used here.  At low-$Q$ the Compton-shift is small and the
Compton and elastic signals cannot be explicitly separated unless a higher
energy resolution measurement is made, for example, using an analyzer crystal.
However, the Compton intensity is much lower and the coherent scattering 
intensity is much larger.  In this region a theoretically calculated 
Compton signal can be subtracted from the data containing both elastic and
Compton scattering.   Uncertainties in this process have a very small 
effect on the resulting $S(Q)$.
Fig.~\ref{fig;fig3} shows the signals from the Compton and elastic
scattering in the $\rm In_{0.5}Ga_{0.5}As$ sample.  At low $Q$, some
contamination from the elastic scattering is apparent in the Compton
channel. For the Compton scattering correction, we followed two steps.
In the high $Q$ region the Compton scattered signal was directly
removed by integrating a narrow region of interest in the MCA spectrum
which only contained the elastic peak. In the low $Q$ region we
calculated the theoretical Compton
scattering~\cite{thijs;jac84,table;volc95} and subtracted this from
the combined (unresolved) Compton plus elastic scattering
signal. These two regions were smoothly interpolated using a window
function, following the method of Ruland~\cite{rulan;bjap64} in which
the theoretical Compton intensity is smoothly attenuated with
increasing~$Q$.  

At very high $Q$ values, due to the Debye-Waller factor, the Bragg peaks
in the elastic scattering signal disappear and the normalized intensity
asymptotes to $\langle f(Q)^2\rangle$.  This fact allows us to obtain an
absolute data normalization by scaling the data to line up with
$\langle f(Q)^2\rangle$ in the high-$Q$ region of the diffraction pattern.
Finally, the total scattering structure
function, $S(Q)$, is then obtained using Eq.~\ref{eq;soq} and the
corresponding PDFs, $G(r)$ are obtained according to Eq.~\ref{eq;gr}.


\section{Results}

\begin{figure}[!tb]
  \centering \includegraphics[angle=0,width=3.3in]{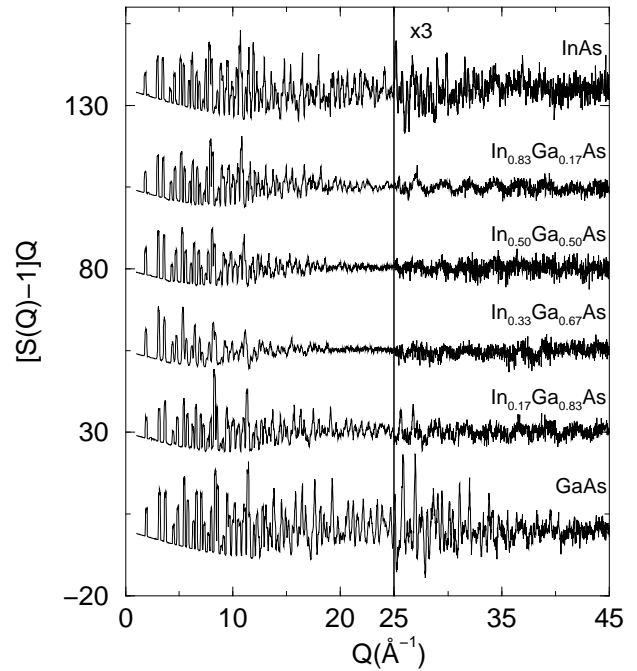}
  \caption{ The reduced total scattering structure function $[S(Q)-1]Q$
  for In$_x$Ga$_{1-x}$As measured at 10K. The data-sets are offset for
  clarity.  The high-$Q$ region is shown on an expanded scale ($\times$ 3) to
  highlight the presence of diffuse scattering.}  
  \label{fig;fig4}
\end{figure}

Fig.~\ref{fig;fig4} shows the experimental
reduced total scattering structure functions, $F(Q)=Q[S(Q)-1]$, for the
$\rm In_{x}Ga_{1-x}As$ alloys measured at 10~K.
It is clear that the Bragg peaks are persistent up to 
$Q \sim 35$ \AA$^{-1}$ in the end-members, GaAs and InAs. 
This reflects both the long range order of the crystalline samples and the
small amount of 
positional disorder (dynamic or static) on the atomic scale. 
In the alloy samples, however, 
the Bragg peaks disappear at much lower $Q$-values but still many sharp 
Bragg peaks are present in the mid-low $Q$ region. 
Instead, oscillating diffuse 
scattering which contains local structural information is evident in high 
$Q$ region. The observation of Bragg peaks reflects the presence of
long-range crystalline order in these alloys.  The fact that the Bragg
peak intensity disappears at lower $Q$-values in the alloys 
than the end-members reflects
that there is significant atomic scale disorder in the alloys as
expected.  The oscillating diffuse scattering in the high-$Q$ region originates
from the stiff nearest-neighbor In-As and Ga-As covalent bonds.  

\begin{figure}[!tb]
 \centering \includegraphics[angle=0,width=3.3in]{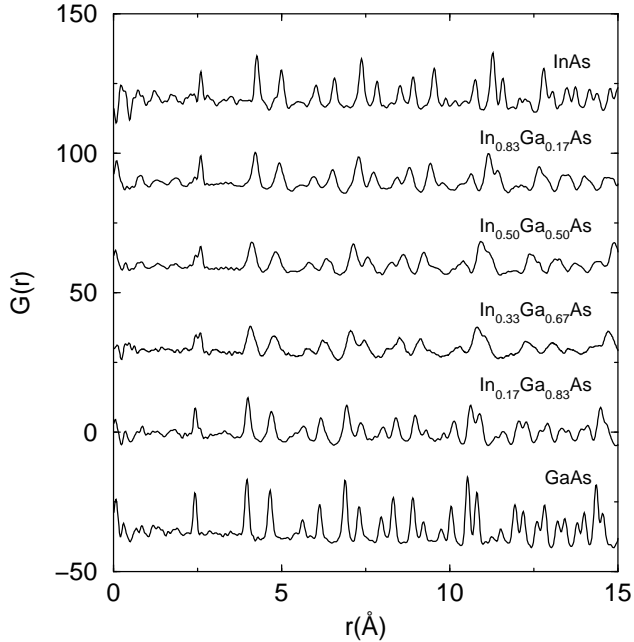}
 \caption{ The reduced PDF, $G(r)$ for In$_x$Ga$_{1-x}$As
            measured at 10~K. The data-sets are offset for clarity.}
  \protect
  \label{fig;fig5}
\end{figure}

Fig.~\ref{fig;fig5} shows the corresponding reduced PDFs, $G(r)$, obtained 
using Eq.~\ref{eq;gr}. 
In the alloys, it is clear that the first peak is split into a doublet
corresponding to shorter Ga-As and longer In-As
bonds.~\cite{petko;jap00} The position in $r$ of the left and right
peaks does not disperse significantly on traversing the alloy
series. This shows that the local bond lengths stay close to their
end-member values and do not follow Vegard's law, in agreement with the
earlier XAFS\cite{mikke;prl82} and PDF\cite{petko;prl99}
reports. However, already by 10 \AA\ the structure is behaving much
more like the average structure.  For example, the doublet of PDF
peaks around 11 \AA\ in GaAs (Fig.~\ref{fig;fig5}) remains a doublet (it
doesn't become a quadruplet in the alloys) and disperses smoothly
across the alloy series to its position at around 12 \AA\ in the pure
InAs.  This shows that already by 10~\AA\, the structure is
exhibiting Vegard's law type behavior.

It is also notable that for the nearest neighbor PDF peak, the peak widths 
are almost the same 
in both alloys and end-members but for the higher neighbors, the peaks 
are much broader in the alloys than in the end-members.


\section{Atomic displacements in the alloys}

\subsection{Modeling}\label{sec;cluster}

A simplified view of the structural disorder in the $\rm A_xB_{1-x}C$
type tetrahedral alloys can be intuitively visualized by considering
simple tetrahedral clusters centered about C sites (the unalloyed
site).  In the random alloy this site can have 4 A-neighbors (type-I),
3 A- and 1 B-neighbors (II), 2 A- and 2 B-neighbors (III), 1 A and 3
B-neighbors (IV) or 4 B neighbors (V). We assume that the mixed site
(A,B) atoms stay on their ideal crystallographic positions.  By
considering each cluster type in turn we can predict the qualitative nature
of the atomic displacements present in the alloy. 
Let the A atoms be larger than the B atoms.
\begin{figure}[!tb]
  \centering \includegraphics[angle=0,width=3.3in]{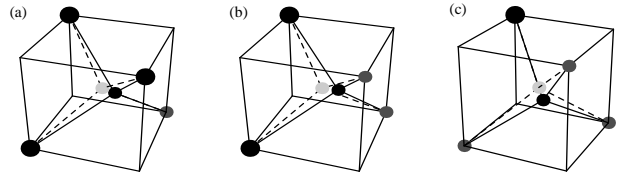}
  \caption{ Schematic diagram of As displacements in cluster (a) type II, 
       (b) type III, and (c) type IV. Cluster types are discussed in the text.
        At the corner, large dark circle and small grey circle show In and
        Ga atom position respectively. At the center, the grey and dark 
        circles correspond to the As atom position before 
        and after displacement, respectively.}
  \protect
  \label{fig;fig6}
\end{figure} 
In clusters of type I and V the C atom will not
be displaced away from the center of the tetrahedron.
As is shown in Fig.~\ref{fig;fig6}, in type II clusters the C atom will
displace away from the center directly towards the B atom.  
This is a displacement in a $\langle
111\rangle$ crystallographic direction. In type III clusters
it will displace in a direction
between the two B atoms along a $\langle 100\rangle$ crystallographic
direction. Finally, in
type IV clusters it will again be a $\langle 111\rangle$ type
displacement but this time in a direction directly away from the
neighboring A atom.
Such a cluster model was used to make quantitative comparisons
with the nearest neighbor 
bond distances observed in XAFS measurements\cite{mikke;prl82} over the 
whole alloy series.\cite{balza;prb85}  However, it was later shown
that the agreement was largely accidental due to the boundary conditions
chosen that all the clusters should have the average size determined
from Vegard's law.\cite{marti;prb84}  Nonetheless, it is interesting to compare
the prediction of this simple cluster model with the nearest-neighbor
PDF peaks measured here since, for the first time, we have an accurate
measurement of the bond length {\it distributions} as well as the bond lengths
themselves.
 
\begin{figure}[!tb]
  \centering \includegraphics[angle=0,width=3.3in]{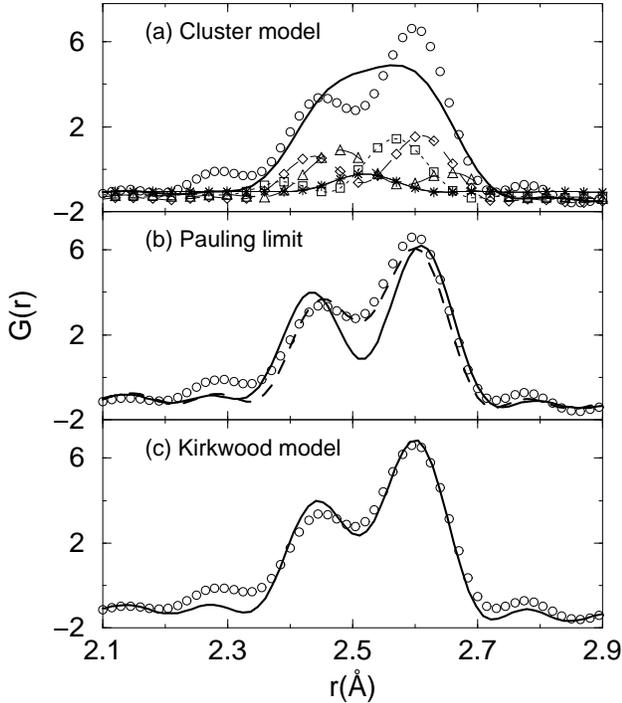}
  \caption{ Comparison between Experimental PDF (open circles) 
   and model PDF (solid line) for $\rm In_{0.5}Ga_{0.5}As$.
   (a) Tetrahedral cluster model with no disorder present 
   on (In,Ga) sublattice. 
   The sub-peaks represent the contributions from each type of cluster. 
   Type I ($\times$), type II ($\Box$), type III ($\Diamond$), 
   type IV ($\triangle$), and type V($\ast$). 
   (b) The model PDF is calculated in the Pauling limit. The peak positions
    were obtained from the InAs and GaAs bond lengths in the 
    end-members (solid line) and the InAs and GaAs bond lengths in 
    the $\rm In_{0.5}Ga_{0.5}As$ PDF (dashed line). 
    See the text for details. 
    (c) Kirkwood supercell model. 
   }
  \protect
  \label{fig;fig7}
\end{figure}

Each cluster is independently relaxed according to the prescription of
Balzarotti~{\it et al.}~\cite{balza;prb85} to get the bond-lengths
within each cluster type.  Assuming a random alloy the number of each
type of cluster that is present can be estimated using a binomial
distribution.  This gives the static distribution of bond lengths
predicted by the model.  These are then convoluted with the broadening
expected due to thermal motion.  This was determined by
measuring the width of the nearest neighbor peaks in the end-member
compounds, InAs and GaAs.  The result is shown in 
Fig.~\ref{fig;fig7}(a).  It is clear that, although the
cluster model gets the peak {\it positions} reasonably correct as
exemplified by the agreement it gets with XAFS
data,~\cite{balza;prb85} it does rather a poor job of explaining the
{\it shape} of the measured pair distribution.  The major discrepancy
is that too much intensity resides at, or close to, the undisplaced
position leading to an unresolved broad first PDF peak in sharp
contrast to the measurement.  In contrast, we show in
Fig.~\ref{fig;fig7}(b) the nearest neighbor atomic pair
distribution in the Pauling limit,~\cite{pauli;bk67} again broadened
by thermal motion.  The peak positions were obtained by using the
bond-lengths of the end-member compounds.  It is clear that this
actually does a better job than the cluster model, though it slightly,
and not surprisingly, overemphasizes the splitting.  The dashed line in this
figure shows the peak profile that we obtain if we make the
assumption that the nearest neighbor bond length changes in the alloy
as seen in the Z-plot,~\cite{mikke;prl82,petko;prl99} but there is no 
increase in the bond length distribution. 
Again, this gives rather good agreement emphasizing the fact that there is 
very little inhomogeneous strain to the covalent bond length due to the 
alloying.~\cite{petko;prl99}

\begin{figure}[!tb]
  \centering \includegraphics[angle=0,width=3.3in]{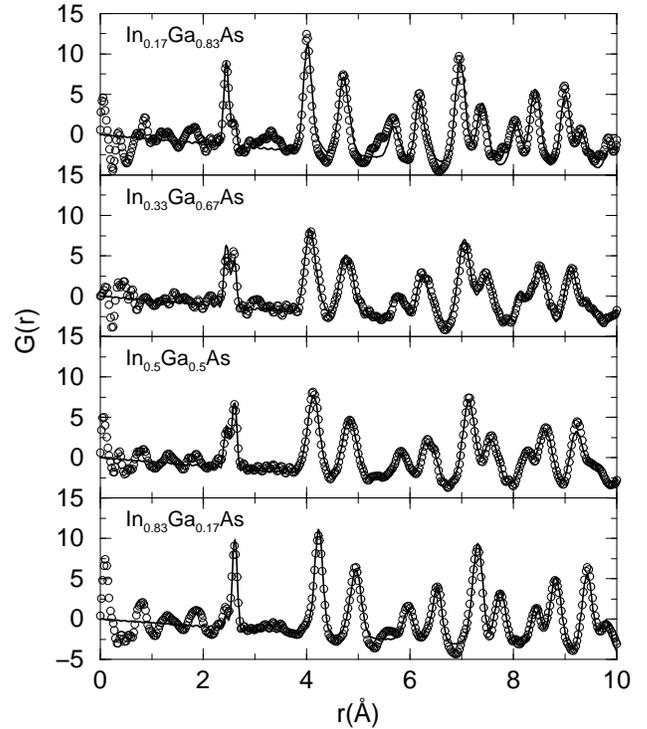}
  \caption{ Comparison between Experimental PDF (open circles) 
   and model PDF (solid line) for $\rm In_{x}Ga_{1-x}As$.
   The model was the Kirkwood supercell model. The parameters $\alpha$ and 
   $\beta$ are refined from the end-members and the PDFs for the alloys
   shown here are then calculated with no adjustable parameters.
   }
  \protect
  \label{fig;fig8}
\end{figure}

A better model for the structure of these
alloys~\cite{petko;prl99,chung;prb99} is obtained from a relaxed
supercell of the alloy system using a Kirkwood
potential.~\cite{kirkw;jcp39} The potential contains nearest neighbor bond
stretching force constants $\alpha $ and force constants $\beta $ that
couple to the change in the angle between adjacent nearest neighbor bonds.
In this relaxed supercell model, the
force constants were adjusted to fit the end-members~\cite{cai;prb92ii} with 
 $\alpha _{{\rm {%
Ga-As}}}$ = 96N/m, $\alpha _{{\rm {In-As}}}$ = 97N/m, $\beta _{{\rm {Ga-As-Ga%
}}}$ = $\beta _{{\rm {As-Ga-As}}}$ = 10N/m and $\beta _{{\rm {In-As-In}}}$ = 
$\beta _{{\rm {As-In-As}}}$ = 6N/m. 
The additional angular force constants
required in the alloy are taken to be the geometrical mean, so that $\beta _{%
{\rm {Ga-As-In}}}$ = $\sqrt{(\beta _{{\rm {Ga-As-Ga}}}.\beta _{{\rm {In-As-In%
}}})}$. 
The PDFs for the alloys could then be calculated
in a self-consistent way for all the alloys with no adjustable
parameters.~\cite{chung;prb99} In this model, the lattice dynamics are
also included in a completely self-consistent way.  Starting with the
force constants and the Kirkwood potential, the thermal broadening of
the PDF peaks at any temperature can be determined directly from the
dynamical matrix and this is how the PDFs were calculated in the
present case.\cite{petko;prl99} The model-PDF is plotted with the data
in Fig.~\ref{fig;fig8} with the nearest-neighbor peak shown on an
expanded scale in Fig.~\ref{fig;fig7}(c).  
The excellent agreement with the data over the
entire alloy range suggests that the simple Kirkwood potential
provides an adequate starting point for calculating distorted alloy
structures in these III-V alloys.
Note that in comparing with experiment, the theoretical PDF has 
been convoluted with a Sinc function to incorporate the truncation of the experimental data at $Q_{max} = 45$~\AA . 

\subsection{3-D atomic probability distribution}

\begin{figure}[!tb]
  \centering \includegraphics[angle=0,width=3.3in]{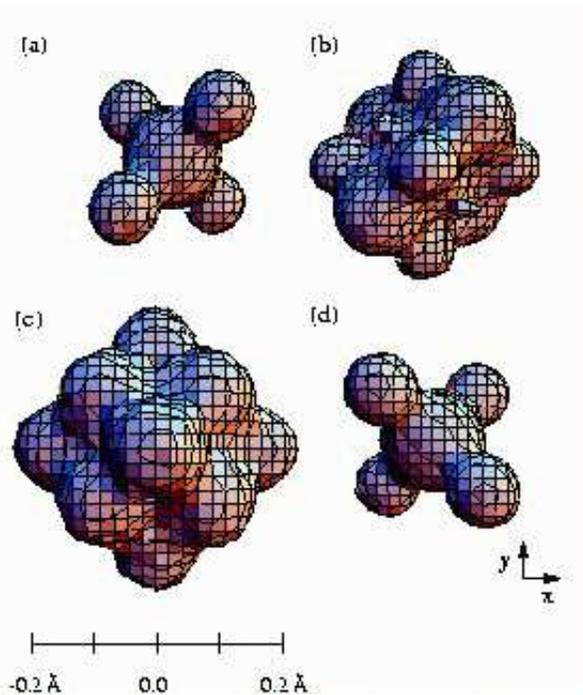}
  \caption{ Iso-probability surface for the ensemble averaged As atom
  distribution.  The surfaces plotted all enclose the volume where As
  atoms will be found with 68~\% probability. (a) $\rm
  In_{0.17}Ga_{0.83}As$ (b) $\rm In_{0.33}Ga_{0.67}As$ (c) $\rm
  In_{0.50}Ga_{0.50}As$ (d) $\rm In_{0.83}Ga_{0.17}As$. In each case,
  the probability distribution is viewed down the [001] axis.}
  \protect \label{fig;fig9}
\end{figure}

\begin{figure}[!tb]
  \centering \includegraphics[angle=0,width=3.3in]{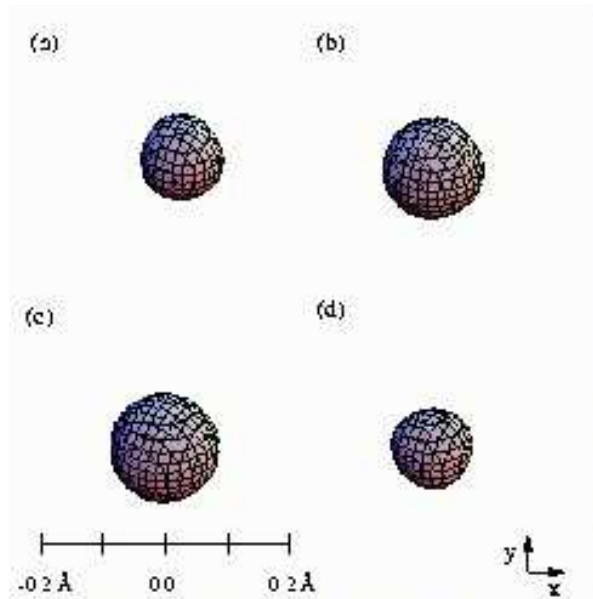}
  \caption{ Iso-probability surface for the ensemble averaged (In,Ga) atom
  distribution.  The surfaces plotted all enclose the volume where As
  atoms will be found with 68~\% probability. (a) $\rm
  In_{0.17}Ga_{0.83}As$ (b) $\rm In_{0.33}Ga_{0.67}As$ (c) $\rm
  In_{0.50}Ga_{0.50}As$ (d) $\rm In_{0.83}Ga_{0.17}As$. In each case,
  the probability distribution is viewed down the [001] axis.
  These surfaces are plotted on the same scale as those in Fig.~9.}
  \protect \label{fig;fig10}
\end{figure}

Now, we analyze the relaxed supercell of alloy system obtained 
using a Kirkwood 
potential to get the average three
dimensional atomic probability distribution of As and (In,Ga) atoms.
Fig.~\ref{fig;fig9} shows iso-probability surfaces for the As 
site in the $\rm In_{x}Ga_{1-x}As$ alloy. 
The probability distributions were created
by translating atomic positions of the displaced arsenic atoms in the
supercell ($20\times20\times20$ cubic cell) into a single unit cell.  
To improve statistics, this was done 70 times. 
The surfaces shown enclose a volume where the As atom will be found
with 68~\% probability.
The probability distribution is viewed down the [001] axis.
It is clear that the As atom displacements, though highly symmetric, are
far from being isotropic. 
The same procedure has been carried out to elucidate the atomic probability 
distribution on the (In,Ga) sublattice.  The results are shown in 
Fig.~\ref{fig;fig10}, plotted on the same scale as in
Fig.~\ref{fig;fig9}.
In contrast to the As atom static distribution, the (In,Ga) probability
distribution is much more isotropic and sharply peaked in space around
the virtual crystal lattice site. 

In all compositions, the As atom distribution is highly anisotropic as evident
in Fig.~\ref{fig;fig9} with large displacements along
$\langle 100\rangle$ and $\langle 111\rangle$ directions. This can be
understood easily within cluster model as we discussed in
Section~\ref{sec;cluster}. The $\langle 100\rangle$ displacements
occur in type III clusters and the $\langle 111\rangle$
displacements occur in type II and IV clusters. 
This also explains why, in the gallium rich alloy in which
the three and four Ga cluster is dominant, the major As atom
displacements are along [111], [1$\bar{1}\bar{1}$],
[$\bar{1}$1$\bar{1}$], and [$\bar{1}\bar{1}$1] as we observed in
Fig.~\ref{fig;fig9}(a). On the contrary, in the indium rich
alloy, the major displacements are along [$\bar{1}\bar{1}\bar{1}$],
[$\bar{1}$11], [1$\bar{1}$1], and [11$\bar{1}$], as can be clearly seen
in Fig.~\ref{fig;fig9}(d).

The atomic probability distribution obtained from the Kirkwood model
for the (In,Ga) sublattice is shown in
Fig.~\ref{fig;fig10}.  As we discussed, this is much more
isotropic (though not perfectly so), and more sharply peaked than the
As atom distribution.  However, contrary to earlier
predictions,~\cite{balza;prb85} and borne out quantitatively by the
supercell modeling, {\it there is significant static disorder
associated with the (In,Ga) sublattice}. 
In order to compare the magnitude of the static distortion of the (In,Ga) 
sublattice with that of the As sublattice, we calculated the standard 
deviation, $\sigma$, of the As and (In,Ga) atomic
probability distributions. This was calculated using  
$\sigma_i = \sqrt{{1\over {N-1}} \sum_{k=1}^N \,(d_i(k))^2}, (i={x,y,z}) $,
where $d_i$ refers to the displacement from the undistorted sublattice 
of atoms in the model supercell in
$x$, $y$, and $z$ directions, and $N$ is the total number of atoms in 
the supercell. 
Table \ref{tab;stdev} summarizes the
values of $\sigma$ for the As and (In,Ga) atomic
probability distributions in the alloys. 
\begin{table}[!tb]
  \centering
  \caption{Standard deviation of the As and (In,Ga) atom distributions in 
           In$_{x}$Ga$_{1-x}$As alloys obtained from the Kirkwood model. 
           The numbers in parentheses are the estimated 
	   error on the last digit. 
           For both As, and 
          (In,Ga) atoms, $\sigma=\sigma_x= \sigma_y= \sigma_z $.
	   See text for details.}

  \label{tab;stdev}
  \begin{tabular}{|c|c|c|c|c|}
                    &  x=0.17  &  x=0.33   &    x=0.50   &  x=0.83   \\
  \hline    
    $\sigma(As)$ (\AA)     &  0.072(1) & 0.092(1) & 0.097(1) & 0.074(1) \\ 
    $\sigma(In,Ga)$ (\AA)&  0.044(1) & 0.058(1) & 0.060(1) & 0.048(1) \\
    ${\sigma(In,Ga)\over \sigma(As)}$ & 0.61 & 0.63   & 0.62 & 0.61 \\
\end{tabular}
\end{table}
It shows that for all compositions
the static disorder on the (In,Ga) sublattice is around 60\% of the 
disorder on the As sublattice. 
These static distortions give rise to a broadening of PDF peaks as described
in Ref.~27 and evident in Fig.~\ref{fig;fig5} of this paper.
To evaluate the static contribution to the PDF peak broadening, $\sigma_D$, 
from the $\sigma's$ reported in Table  \ref{tab;stdev} 
we used the following expression:
\begin{equation}
  \sigma_D^2 =  \sigma_a^2 +  \sigma_b^2,  
  \label{eq;msqdis}
\end{equation}
where $a$, $b$ can be As, or (In,Ga).
For example, for $x$ = 0.5 alloy, we get 
$\sigma_{D}^2$ = 0.0188(4) \AA$^2$ for As-As peaks in the PDF, 
0.0130(4) \AA$^2$ for As-(In,Ga) peaks and
0.0072(3) \AA$^2$ for (In,Ga)-(In,Ga) peaks. 
These values are in good agreement with the
mean square static PDF peak broadening of As-As, As-(In,Ga) and 
(In,Ga)-(In,Ga) peaks, shown in Fig.~4 of Ref.~27, of 
0.0187(1) \AA$^2$, 0.0128(1) \AA$^2$, and 0.0053(1) \AA$^2$
respectively.

\section{Correlated atomic displacements}

We have shown that on the average, atomic displacements of As atoms in
$\rm In_{x}Ga_{1-x}As$ alloy are highly directional. In this section,
we would like to address the question whether these atomic
displacements are correlated from site to site.  To investigate this
we have calculated theoretically the diffuse scattering intensity which 
would be obtained
from the relaxed Kirkwood supercell model and compare it with the
known experimental diffuse scattering.

\begin{figure}[!tb]
  \centering \includegraphics[angle=-90,width=3.3in]{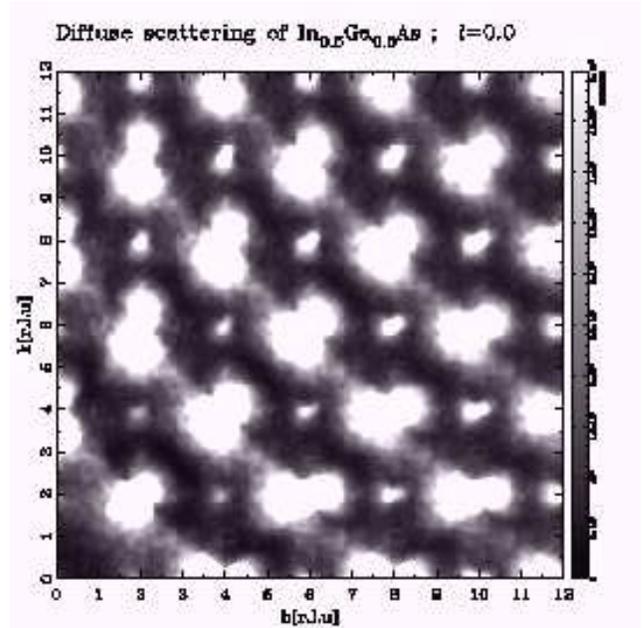}
  \caption{ Single crystal diffuse scattering intensity obtained from the
   relaxed supercell model for the $\rm In_{0.5}Ga_{0.5}As$ alloy. 
   The cut shown is the
   diffuse intensity expected in the $(hk0)$ plane of reciprocal space.
   Bragg peaks have been removed for clarity. See text for details.}
  \protect
  \label{fig;fig11}
\end{figure}

The Fig.~\ref{fig;fig11} shows diffuse scattering of
$\rm In_{0.5}Ga_{0.5}As$ alloy calculated using the DISCUS
program.\cite{proff;jac97} In this calculation the Bragg-peak
intensities have been removed. Strong diffuse scattering is evident at
the Bragg points in the characteristic butterfly shape pointing
towards the origin of reciprocal space.  This is the Huang scattering
which is peaked close to Bragg-peak positions and has already been
worked out in detail.\cite{barab;jpcm99}

\begin{figure}[!tb]
  \centering \includegraphics[angle=-90,width=3.3in]{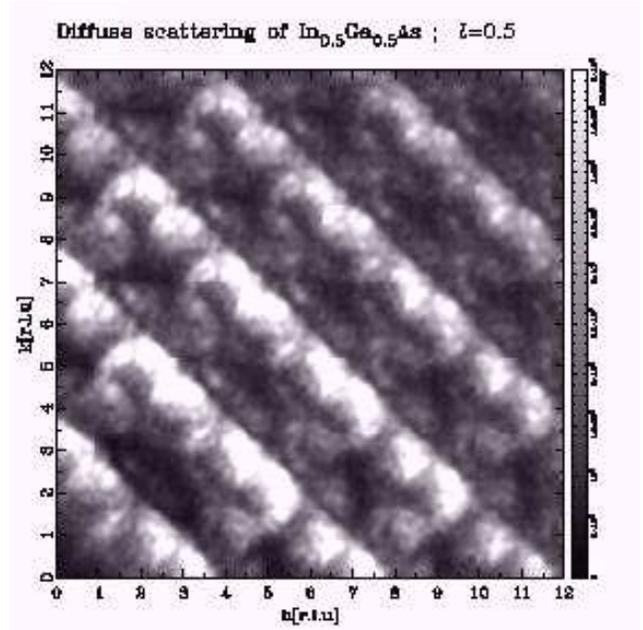}
  \caption{ Single crystal diffuse scattering intensity obtained from the
   relaxed supercell model for the $\rm In_{0.5}Ga_{0.5}As$ alloy. 
   The cut shown is the
   diffuse intensity expected in the $(hk0.5)$ plane of reciprocal space.
   Bragg peaks have been removed for clarity. See text for details.}
  \protect
  \label{fig;fig12}
\end{figure}

In addition to this, clear streaks are apparent running perpendicular
to the [110] direction. The diffuse scattering calculations on $(hkl)$
planes where $l\neq 0,\rm integer$ (Fig.~\ref{fig;fig12}),
show that these
diffuse streaks are extended along the $[00l]$ direction consisting of
sheets of diffuse scattering perpendicular to the [110] direction of
reciprocal space.  Diffuse scattering with exactly this (110) symmetry
was observed in the TEM study of $\rm
In_{0.53}Ga_{0.47}As$.\cite{glas;pmb90} Careful observation of our
calculated diffuse scattering indicates that the diffuse scattering has
a maximum on the low-$Q$ side of the ($hk0$) planes passing through the
Bragg points, with an intensity minimum on the high-$Q$ side of these
planes.  This is characteristic size-effect scattering obtained from
correlated atomic displacements due to a mismatch between chemically
distinct species as recently observed in a single-crystal diffuse
scattering study on Si\sb{1-x}Ge\sb{x},\cite{bollo;xxx00} for example.
This asymmetric scattering was clearly observed in the earlier
diffuse scattering study on $\rm
In_{0.53}Ga_{0.47}As$.\cite{glas;pmb90}

The single-crystal diffuse scattering intensity which is piled up far
from the Bragg-points is giving information about intermediate range
ordering of the atomic displacements.  It is interesting that it is
piled up in planes perpendicular to [110] whereas the local atomic
displacements are predominantly along $\langle 100\rangle$ and
$\langle 111\rangle$ directions.  This observation underscores the
complementarity of single-crystal diffuse scattering and real-space
measurements such as the PDF. The real-space measurements are mostly
sensitive to the {\it direction and magnitude of local atomic
displacements} and less sensitive to how the displacements are
correlated over longer-range (though this information is in the data).
On the other hand, the single crystal diffuse scattering immediately
yields the intermediate range correlations of the displacements but
one has to work harder to extract information about the size and
nature of the local atomic displacements. Used together these two
approaches, together with XAFS, can reveal a great deal of
complementary information about the local structure of disordered
materials.

The single crystal diffuse scattering suggests that atomic
displacements are most strongly correlated (i.e., correlated over the
longest range) along [110] directions although the displacements
themselves occur along $\langle 100\rangle$ and $\langle 111\rangle$
directions.  The reason may be that the zinc-blende crystal is
stiffest along [110] directions because of the elastic anisotropy in
the cubic crystal.  This was shown for the case of InAs and was used
to explain why the 5th peak in the PDF (coming from In-As next
neighbor correlations along [110] direction) was anomalously sharp in
both experiments and calculations.\cite{jeong;jpca99} If the material
is stiffer in this direction, one would expect that strain fields from
displacements will propagate further in these directions than other
directions in the crystal correlating the displacements over longer
range. These are consistent with the displacement pair correlation
function calculation by Glas~\cite{glas;prb95} which shows that the
correlation along $\langle 110\rangle$ directions is larger than
correlations along $\langle 100\rangle$ and $\langle 111\rangle$ and
extends further.


\section{conclusions}

In conclusion, we have obtained high real-space resolution PDFs of 
$\rm In_xGa_{1-x}As$ ($0\leq x \leq 1$) alloys using high energy synchrotron 
x-ray diffraction. For this purpose, we developed a data analysis technique 
adequate for high energy synchrotron x-ray diffraction.
The PDFs show a clearly resolved doublet corresponding to the Ga-As and In-As
bond lengths in the first peak of the alloys. Far-neighbors peaks 
are much broader in the alloys than that of the pure end members.

We show that As atom displacements are highly directional and can be
represented as a combination of $\langle 100\rangle$ and $\langle
111\rangle$ displacements. On the contrary, the (In,Ga) atomic
distribution is much more isotropic. 
The magnitude of (In,Ga) sublattice disorder is less than, but 
rather comparable ($\sigma_{(In,Ga)} \sim 0.6\sigma_{As}$) to, 
the As sublattice disorder. 
Also, the single crystal diffuse scattering
shows that atomic displacements are correlated over the longest range
in [110] directions although the displacements themselves occur along
$\langle 100\rangle$ and $\langle 111\rangle$ directions.

All of the available data, including previous XAFS studies,\cite{mikke;prl82}
the present data,\cite{petko;prl99} differential PDF 
data,\cite{petko;jap00} and diffuse scattering
on a closely related system\cite{glas;pmb90} are well explained by 
a relaxed supercell model based on the Kirkwood potential.\cite{chung;prb99}
This study also underscores the importance of having data from
complementary techniques when studying the detailed structure of
crystals with significant disorder.

\acknowledgements { 
We gratefully acknowledge M. F. Thorpe and J. S. Chung for making their 
supercell calculation program available and giving valuable help.
We would like to acknowledge Th. Proffen for discussions about
diffuse scattering in $\rm In_{x}Ga_{1-x}As$ alloys.
This work was supported by DOE through grant DE FG02 97ER45651.
CHESS is supported by the National Science
Foundation through grant DMR97-13424

}


\end{document}